\documentclass[acmsmall]{acmart}
\AtBeginDocument{%
  }

\setcopyright{acmlicensed}
\copyrightyear{2018}
\acmYear{2018}
\acmDOI{XXXXXXX.XXXXXXX}

\acmJournal{JACM}
\acmVolume{0}
\acmNumber{0}
\acmArticle{0}
\acmMonth{0}

\usepackage{hyperref}
\usepackage{algorithm}
\usepackage{algorithmicx}
\usepackage{algpseudocode}
\usepackage{graphicx}   %
\usepackage{subcaption} %

\newcommand{\approach}{PromptSE}
\newcommand{\lightapproach}{PromptSELight}

\begin{document}

\title{Prompt Stability in Code LLMs: Measuring Sensitivity across Emotion- and Personality-Driven Variations}

\author{Wei Ma}
\email{weima@smu.edu.sg}
\affiliation{%
  \institution{Singapore Management University}
  \country{Singapore}
}

\author{Yixiao Yang }
\affiliation{%
  \institution{Capital Normal University}
  \city{Beijing}
  \country{China}
  }
\email{}

\author{Jingquan Ge}
\authornote{corresponding author}
\affiliation{%
  \institution{Nanyang Technological University}
  \city{}
  \country{Singapore}
}
\email{jingquan.ge@ntu.edu.sg}

\author{Xiaofei Xie }
\affiliation{%
	\institution{Singapore Management University}
	\country{Singapore}
}

\author{Lingxiao Jiang}
\affiliation{%
	\institution{Singapore Management University}
	\country{Singapore}
}

\renewcommand{\shortauthors}{W. Ma et al.}

\begin{abstract}
Code generation models are widely used in software development, yet their sensitivity to prompt phrasing remains under-examined. Identical requirements expressed with different emotions or communication styles can yield divergent outputs, while most benchmarks emphasize only peak performance. We present \approach{} (Prompt Sensitivity Evaluation), a framework that creates semantically equivalent prompt variants with emotion and personality templates, and that evaluates stability using probability-aware continuous scoring, or using binary pass rates when logits are unavailable. The results are aggregated into a proposed area-under-curve metric (AUC-E) for cross-model comparison. Across 14 models from three families (Llama, Qwen and Deepseek), our study shows that performance and stability behave as largely decoupled optimization objectives, and it reveals architectural and scale-related patterns that challenge common assumptions about model robustness. The framework supports rapid screening for closed-source models as well as detailed stability analysis in research settings. \approach{} enables practitioners to quantify performance–stability trade-offs for deployment and model selection, positioning prompt stability as a complementary evaluation dimension alongside performance and fairness, and contributing to more trustworthy AI-assisted software development tools.
\end{abstract}

\ccsdesc[500]{Software and its engineering~Software notations and tools}
\ccsdesc[300]{Computing methodologies~Natural language processing}
\ccsdesc[100]{Information systems~Evaluation of retrieval results}
\ccsdesc[100]{Information systems~Evaluation methodologies}

\keywords{Code LLMs, prompt stability, software engineering, evaluation, robustness}

\maketitle

\section{Introduction}

Code generation models have become integral to software development but they exhibit a critical challege: \textit{prompt sensitivity}, substantial performance variations from semantically equivalent but differently phrased inputs. Consider a frustrated developer at 2 AM asking ``I'm stuck on this recursive problem and need to find the longest increasing subsequence, any ideas?'' versus a formal request ``Implement a function that computes the length of the longest increasing subsequence in an array''. While both seek identical functionality, the emotional context and communication style can trigger dramatically different model behaviors~\cite{zhang-etal-2024-codefort,cao2024worst}, with performance swings up to 40\% reported for production systems~\cite{he2024prompt}. This instability undermines deployment reliability, evaluation comparability, and user trust.

This instability exposes important limitations in current evaluation approaches. While code generation evaluation has evolved to execution-based metrics like Pass@k~\cite{chen2021evaluating}, these methods capture only functional correctness. Recent  works~\cite{cao2024worst,guo2025personality,razavi2025promptset,chen2024nlperturbator} have begun studying prompt sensitivity from different angles. Cao et al.~\cite{cao2024worst} demonstrate that even LLMs experience over 45\% accuracy swings between best and worst prompt phrasings for identical tasks, highlighting the severity of this issue. 
Chen et al.~\cite{chen2024nlperturbator} studies how the natural perturbations affect code generation such as typo while Guo et al.~\cite{guo2025personality} show that personality-aligned prompts can improve code generation pass rates, revealing how psychological factors influence model behavior.
Razavi et al.~\cite{razavi2025promptset} introduce benchmarks for evaluating prompt sensitivity across various NLP tasks, though their focus remains on documenting performance variations rather than establishing unified stability metrics.

Existing studies examine prompt sensitivity through performance gains or losses, but none provide a principled framework to evaluate stability under natural expression variations such as emotional or stylistic changes. Practitioners therefore lack systematic tools to assess this fundamental reliability dimension in code generation models.
We view prompt stability from a different perspective: \textit{\textbf{how can we systematically quantify and compare stability as a measurable model property?}} Unlike Chen et al.~\cite{chen2024nlperturbator} who analyze performance drops, or Guo et al.~\cite{guo2025personality} who optimize for performance gains, we focus on establishing stability as a quantifiable evaluation dimension. We propose \textbf{\textit{\approach{}}} (Prompt Sensitivity Evaluation), introducing an algorithm that transforms stability assessment into standardized, comparable metrics through our novel measurement methodology.

We evaluate 14 models across three architecture families (Llama, Qwen, DeepSeek) using HumanEval~\cite{chen2021evaluating} with semantically equivalent variants (14,760 in total). Each variant prompt generates 16 samples under uniform decoding, with statistical analysis employing robust methods including correlation tests, confidence intervals~(CI), and false discovery rate (FDR)~\cite{benjamini1995controlling} correction. Our study includes four research questions: (1) the relationship between performance and stability; (2) how perturbation magnitude affects sensitivity across model sizes and families; (3) whether \lightapproach{} approximates \approach{}; (4) how emotional factors affect performance and calibration.
Our findings challenge conventional assumptions about code generation models. First, performance and stability are decoupled: models distribute across all four quadrants of the performance-stability space (Spearman $\rho = -0.433$, $p = 0.122$), enabling practitioners to optimize for specific requirements without automatic tradeoffs. Second, prompt stability exhibits non-monotonic scaling patterns, with smaller models (e.g., Qwen-1.5B) achieving superior stability (AUC-E $0.646$) compared to larger models, suggesting stability requires explicit optimization beyond scale. Third, emotional prompting reveals model-specific vulnerabilities and confidence miscalibration patterns not captured by traditional benchmarks. Finally, our dual-pathway evaluation achieves reasonable consistency (\lightapproach{} vs \approach{}: Pearson is about $0.72$), enabling both rapid screening and detailed analysis for deployment decisions.

Our contributions advance code generation evaluation in three key dimensions:
\begin{enumerate}
 \item \textit{Methodological innovation}: We introduce the first systematic framework for quantifying prompt stability, moving beyond performance measurement to establish stability as a distinct evaluation dimension. Our psychologically-grounded perturbation templates capture simulated developer communication patterns through emotional and stylistic variations.
    \item \textit{Technical advancement}: We develop \approach{}, a probability-aware continuous evaluation metric that distinguishes high-confidence solutions from lucky guesses, and AUC-E, a standardized 0-1 stability measure enabling systematic cross-model comparisons. Our unified algorithmic framework seamlessly accommodates both open-source and closed-source models through dual evaluation pathways.
    \item \textit{Empirical insights}: We provide the first comprehensive stability landscape across 14 models and three major model-architecture families, revealing that performance and stability are decoupled optimization objectives and establishing reproducible protocols that challenge conventional scaling assumptions while informing practical deployment decisions.
\end{enumerate}

This work establishes prompt stability as a measurable construct for evaluating code generation models under natural expression variations, providing a unified framework for cross-model comparisons and deployment decisions.

\section{Methodological Framework}
\label{sec:method_framework}

We introduce \textit{\textbf{\approach{}} (Prompt Sensitivity Evaluation)}, a systematic framework for quantifying code generation model stability under prompt variations. Our core insight is to transform stability from an intuitive concept into a measurable model property through psychologically-grounded perturbations and probability-aware evaluation. Our approach generates semantically equivalent prompt variants through emotion and personality templates, then evaluates model sensitivity using probability-aware continuous scoring (\approach{}) or binary evaluation (\lightapproach{}) depending on model accessibility.
The framework operates under strict semantic and interface invariance constraints, ensuring that all variants preserve computational requirements (e.g., preserving input-output constraints, and complexity bounds) and functional specifications. We employ structured psychological templates to systematically simulate how developers might express identical requirements under different emotional states and personality traits, enabling controlled stability assessment across diverse communication styles.
Our evaluation pipeline computes elasticity curves across three perturbation distances (0.1, 0.2, 0.3), then aggregates them using proposed AUC-E for cross-model comparisons. This dual-pathway design ensures compatibility with both open-source models (with probability access) and closed-source APIs (binary-only evaluation). The framework consists of three core components:

\begin{enumerate}
\item \textit{Emotion-aware variant generator}: Creates semantics-preserving prompt variants using psychological templates (emotion, personality) and perturbation strength controls.
\item \textit{Sensitivity evaluation}: Implements \approach{} for probability-based analysis and \lightapproach{} for binary evaluation, measuring model sensitivity to prompt variations.
\item \textit{Elasticity quantification}: Integrates sensitivity curves across perturbation distances to compute AUC-E for cross-model stability comparison.
\end{enumerate}

\subsection{Emotion-Aware Template-Based Variant Generation Method}
\label{subsec:emotion_aware_variant_generation}

We adopt a template-constrained design philosophy that translates interpretable psychological factors (emotion, personality) and perturbation strength into unified linguistic constraints for controlled, semantics-preserving rewriting. This approach prioritizes semantic preservation while introducing stylistic variations through controlled template libraries.

\subsubsection{Theoretical Modeling Foundation}
\label{subsubsec:theoretical_modeling}

To evaluate model sensitivity to prompt variants in realistic development contexts, we observe that developers express the same coding requirements differently based on their emotional states (such as frustration or excitement) and personality traits (such as communication style preferences). Contextual factors like task complexity and time pressure further influence how these psychological factors appear in prompt formulation. This natural variation aligns with established research in affective computing, where the valence-arousal model~\cite{russell1980circumplex,pei2024affective,galvao2021predicting,feldman1995variations,schlosberg1952facial} provides a robust basis for mapping emotions to linguistic variations, and personality psychology, which shows that traits like the Big Five dimensions systematically affect language usage~\cite{zhou-etal-2024-evaluating, mairesse-walker-2007-personage, jiang-etal-2023-mpi}.
We model prompt variant generation as a constrained stylistic rewriting problem that preserves task semantics while capturing these natural communication patterns. Our approach makes three key modeling choices:

\begin{itemize}
\item \textbf{Emotion as linguistic style signals}: We treat emotions as short-term style variations using the arousal and valence framework, creating identifiable emotion templates (such as focus, excitement, anxiety) with specific linguistic features and expression patterns.
\item \textbf{Personality as stable baseline}: We represent personality through three dimensions (technical orientation, experience level, collaboration style) that serve as a consistent expression baseline, with emotions layered on top through weighting mechanisms.
\item \textbf{Context-sensitive weighting}: We use task attributes (such as algorithmic complexity, collaboration needs, learning goals) to adjust the influence of emotion and personality factors while maintaining semantic preservation and technical requirements.
\end{itemize}

We formalize variant generation as follows: given an original prompt $p$, emotion template $e$, personality profile $\Pi$, and contextual conditions $x$, we generate a semantically equivalent variant $v$ whose style reflects the constraints from $(e,\Pi,x)$ with rewriting intensity controlled by distance $d$.

\paragraph{Emotion State}
For programming contexts, we define eight core emotion states through structured templates: \textit{focused, excited, confident, tired, calm, anxious, frustrated, and stressed}. Each emotion prompt template consists of three components:
\begin{equation}
\text{Emotion } e = (\text{description}, \text{language characteristics}, \text{expression pattern})
\label{eq:emotion_template}
\end{equation}
where \textit{description} establishes the programmer's role and cognitive state, \textit{language characteristics} defines preferred vocabulary and linguistic features, and \textit{expression pattern} specifies sentence structures and communication styles. This template-based approach follows the valence–arousal tradition in affective computing~\cite{russell1980circumplex, picard1997affective} and enables controllable text generation~\cite{zhou-etal-2018-ecm}.
For example, the ``focused'' state is defined as follows: a programmer in deep focus state with cognitive resources concentrated on technical tasks, using precise and concise technical language with specific terminology and minimal redundancy, and expressing ideas through declarative sentences with strong logic and clear technical direction.

\paragraph{Personality Trait Profile}
We introduce a three-dimensional personality profile
\begin{equation}
\Pi=(T,\,L,\,C),
\label{eq:personality_profile}
\end{equation}
where $T$ denotes the technical orientation, $L$ the experience level, and $C$ the collaboration style. The technical orientation dimension distinguishes between algorithm experts, pragmatic engineers, experimental innovators, and defensive conservatives. The experience level dimension separates junior explorers from senior architects, thereby capturing differences in accumulated expertise. The collaboration style dimension covers logic-driven, collaboration-oriented, plan-systematic, and adaptive-flexible tendencies, which provide complementary perspectives on interpersonal and organizational behaviors in development contexts.

\subsubsection{Template Library Design and Implementation}
\label{subsubsec:template_library_design}

\paragraph{Emotion State Template Library}
We organize eight emotion states: focused, excited, confident, tired, calm, anxious, frustrated, and stressed. Each emotion state follows the template structure defined in Equation~\ref{eq:emotion_template}, providing role setting, linguistic features, and expression patterns for controlled generation~\cite{zhou-etal-2018-ecm, resendiz-klinger-2023-emotion, zhou-etal-2024-evaluating}.

\paragraph{Personality Trait Template Library}
We define personality profiles across three dimensions with distinct linguistic patterns.
\textit{Technical Orientation ($T$)} includes four types: Algorithm Expert (uses theoretical vocabulary like ``complexity'' and ``optimal''), Pragmatic Engineer (emphasizes practical terms like ``build'' and ``deploy''), Experimental Innovator (favors exploratory language like ``explore'' and ``experiment''), and Defensive Conservative (focuses on stability terms like ``ensure'' and ``stable'');
\textit{Experience Level ($L$)} distinguishes Junior Explorers (use concrete expressions and seek guidance) from Senior Architects (employ complex structural language and systems thinking);
\textit{Collaboration Style ($C$)} covers four approaches: Logic-Driven (analytical vocabulary), Collaboration-Oriented (inclusive language), Plan-Systematic (structured expressions), and Adaptive-Flexible (iterative phrasing).

\paragraph{Perturbation Strength Template}
The perturbation distance $d$ controls rewriting intensity across three levels: $d=0.1$ applies light lexical changes, $d=0.2$ introduces moderate style adjustments, and $d=0.3$ creates substantial transformation while preserving semantics.

\subsubsection{End-to-End Emotion-Aware Variant Generation}
\label{subsubsec:variant_generation_algorithm}
Given an original prompt $p$, target emotion $e$, personality profile $\Pi$, and perturbation distance $d$, our variant generation process outputs a candidate set $\mathcal{V} = \{(v, e, \Pi, d)\}$, where each variant $v$ is tagged with its generation parameters for complete traceability. The process operates across multiple distance levels $D = \{0.1, 0.2, 0.3\}$ and generates $K$ candidates per distance layer (we generate 30 variants per original prompt for each distance). 

\paragraph{Template Sampling and Configuration}
The process begins by sampling emotion and personality templates from the predefined libraries based on the specified $(e,\Pi,d)$ configuration. This sampling strategy ensures diverse coverage while maintaining computational efficiency. The emotion templates provide role settings, linguistic features, and expression patterns, while personality templates contribute linguistic mappings across technical orientation, experience level, and collaboration style dimensions.

\paragraph{Prompt Construction and Constraints}
The sampled template elements are translated into natural language generation instructions that guide the rewriting process. The perturbation parameter $d$ controls rewriting intensity, determining how substantially the variant differs from the original prompt. Critical to this stage is maintaining interface invariance: function signatures, type annotations, and import statements remain unchanged. Parameters without default values may be renamed for stylistic variation, while those with defaults must preserve both names and values.

\paragraph{Candidate Generation and Validation}
The system generates multiple diverse candidates using controlled randomness and multi-style resampling. This approach ensures that variants explore different stylistic paths while remaining semantically equivalent to the original prompt. Generated candidates are then validated against structural and interface requirements to ensure they maintain functional equivalence.

\paragraph{Metadata and Storage}
Each validated variant is annotated with its generation parameters (emotion, personality, perturbation distance) and stored in a hierarchical cache organized by dataset, original prompt, and distance layer. This structure enables efficient reuse across different models and facilitates subsequent statistical analyses.

\subsection{Sensitivity Evaluation Methods: \approach{} and \lightapproach{}}
\label{subsec:sensitivity_evaluation}

To quantify model output stability under prompt perturbations and address the core challenge of performance fluctuations caused by natural expression variations, we propose two complementary evaluation methods within a unified direct difference framework. \approach{} provides comprehensive analysis with probability-aware scoring for research scenarios requiring fine-grained insights, while \lightapproach{} offers rapid evaluation suitable for closed-source models and industrial applications where quick stability assessment is essential. This dual-pathway design ensures ecosystem compatibility while maintaining methodological rigor, enabling both deep analysis and practical screening across diverse deployment contexts.
Table~\ref{tab:notation} shows the mathematical notation for our sensitivity evaluation framework.

\begin{table}[]
\centering
\caption{Mathematical Notation and Symbol Definitions}
\label{tab:notation}
\small
\begin{tabular}{ll}
\toprule
\textbf{Symbol} & \textbf{Definition} \\
\midrule
\multicolumn{2}{l}{\textit{Basic Elements}} \\
$p$ & Original prompt from the dataset \\
$d \in \{0.1, 0.2, 0.3\}$ & Perturbation distance levels \\
$V_d^p = \{v_1^{d,p}, v_2^{d,p}, \ldots, v_{n_d}^{d,p}\}$ & Set of variants for prompt $p$ at distance $d$ \\
$m$ & Number of code samples generated per prompt \\
$y_j$ & $j$-th generated code sample ($j \in \{1, 2, \ldots, m\}$) \\
$\pi_j$; $\widehat{\pi}_j$ & Model's probability for the sample $j$; normalized version  \\
\midrule
\multicolumn{2}{l}{\textit{Evaluation Metrics}} \\
$\text{Pass}(p)$ & Pass rate of original prompt $p$ \\
$\text{Pass}(d)$ & Average pass rate of all variants at distance $d$ \\
$\Delta\text{Pass}(d) = |\text{Pass}(p) - \text{Pass}(d)|$ & Pass rate difference \\
$\text{Acc}_{\text{soft}}(p)$ & SoftExec continuous correctness score for prompt $p$ \\
\midrule
\multicolumn{2}{l}{\textit{Aggregation and Final Metrics}} \\
$\text{Elasticity}(p, d)$ & Elasticity of prompt $p$ at distance $d$ \\
$\mathcal{E}(d)$ & Dataset-level average elasticity at distance $d$ \\
$\text{AUC-E}$ & Area Under Curve of Elasticity, our primary sensitivity metric \\
\midrule
\multicolumn{2}{l}{\textit{Implementation Details}} \\
$\mathcal{P}_{\text{prompts}} = \{p_1, p_2, \ldots, p_n\}$ & Prompt dataset \\
$T = \{t_1, t_2, \ldots, t_n\}$ & Corresponding test suites \\
$\mathcal{V}$ & Pre-generated variant cache from Section~\ref{subsubsec:variant_generation_algorithm} \\
$\mathcal{P}$ & Set of all passing code outputs \\
$I[\cdot]$ & Indicator function (1 if test success is true, 0 otherwise) \\
\bottomrule
\end{tabular}
\end{table}

\subsubsection{SoftExec: Probability-Aware Continuous Evaluation}
\label{subsubsec:softexec}

Traditional Pass@k evaluation treats all passing solutions equally, ignoring the model's confidence in its outputs. To capture this probability information, we introduce SoftExec, a continuous evaluation method that weights correctness by the model's generation probability.

\paragraph{Core Concept}
SoftExec computes a weighted correctness score where each generated code sample contributes proportionally to its likelihood under the model:

\begin{equation}
\text{Acc}_{\text{soft}}(p) = \sum_{j=1}^{m} \pi_j \cdot I[y_j \in \mathcal{P}]
\label{eq:softexec_main}
\end{equation}

where $\pi_j$ is the output probability and $I[y_j \in \mathcal{P}]$ indicates whether the sample passes all tests.
To ensure numerical stability and meaningful comparisons, we apply softmax normalization across all $m$ samples from the same prompt:
$
\widehat{\pi}_j = \frac{\pi^{j}}{\sum_{k=1}^{m} \pi^{k}}
\label{eq:softmax_norm}
$, where $\sum_{j=1}^{m} \widehat{\pi}_j = 1$. We use the normalized probability $\widehat{\pi}_j$ instead of $\pi_j$ in practice.

This probability-aware evaluation provides several key benefits. First, it enables confidence-aware scoring by distinguishing between high-confidence correct solutions and lucky guesses, providing a more nuanced understanding of model performance. Second, the method applies normalization procedures that help mitigate numerical instability and reduce the risk of probability collapse, supporting more reliable computations under a wide range of probability values. Finally, SoftExec offers fine-grained sensitivity analysis, supporting the detection of subtle changes in model behavior under prompt perturbations that might otherwise be overlooked by binary evaluation methods.

\subsubsection{\approach{}: Comprehensive Probability-Based Sensitivity Analysis}
\label{subsubsec:pse_plus}

\approach{} provides the most detailed sensitivity analysis by leveraging sequence probabilities from open-source models. The method systematically compares original prompts with their perturbed variants using continuous SoftExec scores.

\paragraph{Unified Evaluation Framework}
Both \approach{} and \lightapproach{} operate within a unified algorithmic framework (Algorithm~\ref{alg:unified_evaluation}) that leverages the pre-generated prompt variants from Section~\ref{subsubsec:variant_generation_algorithm}. The framework employs a mode parameter to seamlessly switch between probability-based continuous evaluation (\approach{}) and binary pass rate evaluation (\lightapproach{}), ensuring methodological consistency while accommodating different computational and model accessibility constraints. This unified approach demonstrates that both methods share identical structural logic, differing only in their scoring mechanisms and sensitivity quantification strategies.

\begin{algorithm}[]
\centering
\caption{Unified Sensitivity Evaluation Process}
\label{alg:unified_evaluation}
    \resizebox{0.9\linewidth}{!}{%
    \begin{minipage}{\linewidth}
\begin{algorithmic}[1]
\Require Prompt set $\mathcal{P}_{\text{prompts}}$, Pre-generated variant cache $\mathcal{V}$, Test suites $T$, Distance levels $D$, Sample count $m$, Evaluation mode $\text{mode} \in \{\approach{}, \lightapproach{}\}$
\Ensure AUC-E score, Elasticity curve $\mathcal{E}(d)$
\State \textbf{Phase 1: Prompt Set Assembly and Evaluation}
\For{each original prompt $p \in \mathcal{P}_{\text{prompts}}$}
    \State $S_p \leftarrow \{p\} \cup \bigcup_{d \in D} \text{load\_cached\_variants}(p, d)$ \Comment{Complete prompt set}
    \For{each prompt $s \in S_p$}
        \If{$\text{mode} = \approach{}$}
            \State $\{y_1^s, \ldots, y_m^s\}, \{\pi_1^s, \ldots, \pi_m^s\} \leftarrow \text{generate\_with\_logprobs}(s, m)$
            \State $\{r_1^s, \ldots, r_m^s\} \leftarrow \text{execute\_tests}(\{y_1^s, \ldots, y_m^s\}, T_p)$
            \State $\text{Score}(s) \leftarrow \text{SoftExec}(\{\pi_j^s\}, \{r_j^s\})$ \Comment{Continuous scoring, $\text{Acc}_{\text{soft}}(s)$}
        \Else \Comment{\lightapproach{} mode}
            \State $\{y_1^s, \ldots, y_m^s\} \leftarrow \text{generate\_code}(s, m)$
            \State $\{r_1^s, \ldots, r_m^s\} \leftarrow \text{execute\_tests}(\{y_1^s, \ldots, y_m^s\}, T_p)$
            \State $\text{Score}(s) \leftarrow \frac{1}{m} \sum_{j=1}^{m} r_j^s$ \Comment{Binary pass rate}
        \EndIf
    \EndFor
\EndFor

\State \textbf{Phase 2: Elasticity Computation}
\For{each original prompt $p \in \mathcal{P}_{\text{prompts}}$}
    \For{each distance level $d \in D$}
        \If{$\text{mode} = \approach{}$}
            \State $\text{differences} \leftarrow \{|\text{Score}(p) - \text{Score}(v)| : v \in V_d^p\}$
            \State $\text{Elasticity}(p, d) \leftarrow 1 - \text{mean}(\text{differences})$
        \Else \Comment{\lightapproach{} mode}
            \State $\text{Score}(d) \leftarrow \frac{1}{|V_d^p|} \sum_{v \in V_d^p} \text{Score}(v)$
            \State $\text{Elasticity}(p, d) \leftarrow 1 - |\text{Score}(p) - \text{Score}(d)|$
        \EndIf
    \EndFor
\EndFor

\State \textbf{Phase 3: Dataset-Level Aggregation}
\For{each distance level $d \in D$}
    \State $\mathcal{E}(d) \leftarrow \frac{1}{|\mathcal{P}_{\text{prompts}}|} \sum_{p \in \mathcal{P}_{\text{prompts}}} \text{Elasticity}(p, d)$
\EndFor
\State $\text{AUC-E} \leftarrow \text{simpson\_integration}(\mathcal{E}, D)$

\State \Return AUC-E, $\mathcal{E}(\cdot)$
\end{algorithmic}
\end{minipage}}
\end{algorithm}

Algorithm~\ref{alg:unified_evaluation} presents a unified framework that encompasses both \approach{} and \lightapproach{} evaluation modes through parameterized execution paths. The algorithm operates in three distinct phases, each designed to handle the core differences between probability-based and binary-based evaluation approaches.

\textbf{Phase 1: Prompt Set Assembly and Evaluation (lines 3-16).} The algorithm begins by constructing a comprehensive evaluation set $S_p$ for each original prompt $p$, which includes both the original prompt and all pre-generated variants across different perturbation distances. The key innovation lies in the conditional evaluation strategy: when operating in \approach{} mode, the algorithm invokes \texttt{generate\_with\_logprobs} to obtain both code samples and their sequence-level log probabilities, subsequently computing continuous SoftExec scores. In contrast, \lightapproach{} mode uses standard \texttt{generate\_code} without probability information, computing binary pass rates directly from test execution results. This dual-path approach ensures that both evaluation modes can leverage the same algorithmic structure while maintaining their distinct computational characteristics.

\textbf{Phase 2: Elasticity Computation (lines 18-27).} The elasticity calculation phase implements mode-specific sensitivity quantification strategies. For \approach{} mode, the algorithm computes individual score differences between the original prompt and each variant at distance $d$, then averages these differences to capture fine-grained sensitivity patterns across all variants. This approach enables detection of subtle variations in model confidence. \lightapproach{} mode, conversely, aggregates variant scores into a distance-level average before computing the absolute difference with the original prompt score. This simplified approach trades granular sensitivity detection for computational efficiency while maintaining mathematical consistency with the direct-difference framework.

\textbf{Phase 3: Dataset-Level Aggregation (lines 29-33).} The final phase standardizes both evaluation modes through identical aggregation procedures. Individual prompt elasticities are averaged across the entire dataset to produce the elasticity curve $\mathcal{E}(d)$, which captures population-level sensitivity patterns. The AUC-E metric is then computed using Simpson's rule numerical integration, providing a single comprehensive measure of overall prompt sensitivity that facilitates cross-model and cross-dataset comparisons.
The unified design maintains methodological consistency across both evaluation modes. In practice, they give broadly comparable AUC-E scores and elasticity curves in many settings. This allows researchers to select the mode that fits their computational constraints and model accessibility with an awareness of these trade-offs.

\paragraph{Elasticity Calculation Principle}
The core insight of \approach{} is that prompt sensitivity can be quantified through the stability of continuous correctness scores. For a given prompt $p$ and distance $d$, elasticity is computed as:

\begin{equation}
\text{Elasticity}(p, d) = 1 - \frac{1}{|V_d^p|} \sum_{v \in V_d^p} |\text{Acc}_{\text{soft}}(p) - \text{Acc}_{\text{soft}}(v)|
\label{eq:pse_plus_elasticity}
\end{equation}

This formulation exhibits several desirable mathematical and interpretive characteristics. The elasticity measure provides intuitive interpretation through its bounded range $\in [0,1]$, where values approaching 1 indicate perfect stability and values near 0 represent maximum sensitivity to prompt perturbations. The formulation assumes a linear relationship between score differences and sensitivity levels, establishing a proportional correspondence that facilitates straightforward sensitivity quantification. Additionally, the method ensures symmetric treatment of variants by giving equal weight to all variants at the same distance level, preventing bias toward particular perturbation patterns and ensuring balanced sensitivity assessment.

\subsubsection{\lightapproach{}: Efficient Binary-Based Sensitivity Analysis}
\label{subsubsec:one_shot}

\lightapproach{} provides a computationally efficient alternative for scenarios where probability information is unavailable (closed-source models) or rapid evaluation is required. It uses binary pass rates instead of continuous scores while maintaining the same direct-difference framework as \approach{}.
This method evaluates the same emotion-aware prompt variants but relies on aggregate pass rate differences rather than probability-based scoring.
The core metric maintains mathematical consistency with \approach{} while operating on discrete pass rates:

\begin{equation}
\text{Elasticity}(p, d) = 1 - |\text{Pass}(p) - \text{Pass}(d)|
\label{eq:one_shot_elasticity}
\end{equation}

where $\text{Pass}(p) = \frac{1}{m} \sum_{j=1}^{m} I[y_j^p \in \mathcal{P}]$ (original prompt pass rate), $\text{Pass}(d) = \frac{1}{|V_d^p|} \sum_{v \in V_d^p} \text{Pass}(v)$ (average variant pass rate) and $\Delta\text{Pass}(d) = |\text{Pass}(p) - \text{Pass}(d)|$ represents the absolute difference in success rates.

\textit{Advantages and Trade-offs.}
\lightapproach{} offers significant practical advantages while accepting certain precision limitations. The method achieves computational efficiency by eliminating expensive log-probability calculations and enabling single-pass sampling with parallel execution. It provides broad applicability as a model-agnostic approach that works with any code generation system, including closed-source models and commercial APIs, making it suitable for industrial applications and rapid prototyping.
However, the binary approach operates at coarser granularity and maynot distinguish between high-confidence and low-confidence correct solutions. This may cause the method to miss subtle sensitivity patterns and exhibit reduced precision when detecting fine-grained behavioral changes in model responses.

\subsection{AUC-E: Area Under Curve of Elasticity}
\label{subsec:auc_e_metric}

To provide a single, interpretable measure of overall prompt sensitivity, we introduce AUC-E (Area Under Curve of Elasticity). This metric aggregates elasticity measurements across all perturbation distances into a unified score, delivering a calibrated 0-to-1 stability scale that enables systematic comparison across models and datasets.
AUC-E captures the intuition that a robust model should maintain consistent performance across a range of prompt perturbations. By integrating the elasticity curve $\mathcal{E}(d)$ over perturbation distances, we obtain a comprehensive sensitivity measure that accounts for both local and global stability patterns.

\textit{Elasticity Curve Construction.}
The elasticity curve is constructed through hierarchical aggregation. First, for each prompt $p$ and distance $d$, compute individual elasticity using the formulations from Equations~\ref{eq:pse_plus_elasticity} and~\ref{eq:one_shot_elasticity}.
For \approach{} mode,
$
\text{Elasticity}(p, d) = 1 - \frac{1}{|V_d^p|} \sum_{v \in V_d^p} |\text{Acc}_{\text{soft}}(p) - \text{Acc}_{\text{soft}}(v)|
$.
For \lightapproach{} mode,
$
\text{Elasticity}(p, d) = 1 - |\text{Pass}(p) - \text{Pass}(d)|
$.
Second, average across all prompts to obtain the elasticity curve:
$
\mathcal{E}(d) = \frac{1}{|\mathcal{P}_{\text{prompts}}|} \sum_{p \in \mathcal{P}_{\text{prompts}}} \text{Elasticity}(p, d)
\label{eq:elasticity_curve}
$.

\textit{AUC-E Computation.}
\label{subsubsec:auc_e_computation}
For our three-distance evaluation setup ($d \in \{0.1, 0.2, 0.3\}$), we compute AUC-E using Simpson's rule~\cite{simpson_numerical_1962}:
$
\text{AUC-E} = \frac{1}{9} \times (\mathcal{E}(0.1) + 4 \times \mathcal{E}(0.2) + \mathcal{E}(0.3))
\label{eq:auc_e}
$.

This integrates the elasticity curve over the interval [0.1, 0.3] and normalizes the result to the [0,1] range, enabling direct comparison across different models and datasets. Higher values indicate better prompt stability, with values approaching 1.0 suggesting highly robust models and values near 0.0 indicating high sensitivity to prompt variations.
The metric can reduce the impact of single-distance outliers and can distinguish between models with similar average performance when their sensitivity patterns differ across perturbation distances.

\section{Evaluation}
\label{sec:evaluation}
\subsection{Research Questions}
To systematically characterize prompt stability in code generation under strict \textit{semantic and interface invariance}, we pursue one overarching goal: derive a probability aware, aggregatable stability measure (SoftExec based elasticity and AUC-E) that enables comparable and reproducible analysis across models. Guided by this goal and our dual evaluation pathways (\approach{} for probability aware scoring and \lightapproach{} for probability free settings), we ask four research questions: (i) how performance and prompt stability are jointly structured across the evaluated models and whether there is evidence of tradeoffs (RQ1); (ii) how \textit{perturbation distance} $d$ shapes sensitivity and to what extent model size and family moderate this effect (RQ2); (iii) whether \lightapproach{} can approximate \approach{} and where the approximation breaks down (RQ3); and (iv) how \textit{valence $\times$ arousal} conditions independently affect correctness and calibration and what mechanism evidence supports these effects (RQ4).

\textbf{RQ1. How are performance and prompt stability jointly structured across the evaluated models?}
Practical deployments balance raw performance (e.g., Pass@k) with insensitivity to prompt phrasing. Rather than presupposing a universal negative relationship, we characterize the joint structure of performance and stability and quantify any association (with uncertainty) at the model level. We construct problem-level elasticity curves from probability-aware continuous correctness (SoftExec), summarize dataset-level stability with AUC-E using Simpson’s rule and normalization, then evaluate correlations and derive a four-quadrant typology, followed by stratified and robustness analyses by model size and family. For models without probability access, we repeat the analysis with \lightapproach{} proxies ($\Delta\text{Pass}$ and \lightapproach{} elasticity) to assess consistency.

\textbf{RQ2. How does perturbation distance shape sensitivity, and to what extent do model size and family moderate this effect across the evaluated models?}
Real usage ranges from light wording tweaks to substantial stylistic shifts. We quantify how perturbation distance $d$ shapes correctness and stability and assess whether model size and family moderate this effect. Under three controlled distance levels $d \in {0.1, 0.2, 0.3}$, we compare the original prompt with its variants to obtain sensitivity curves and aggregated indicators (SoftExec‑based elasticity and AUC‑E). We adopt nonparametric statistics with multiple‑comparison control to estimate main and interaction effects by size and family, visualize pattern differences (e.g., gradual versus irregular), and report effect sizes with uncertainty. When probabilities are unavailable, we complement with discrete proxies ($\Delta\text{Pass}$).

\textbf{RQ3. Can \lightapproach{} approximate \approach{}, and where does it break down?}
Under resource constraints or with closed models, a low cost path is essential. We therefore ask whether \lightapproach{} can recover the rankings and relative comparisons produced by \approach{}, and under what conditions it fails. We align elasticity computed from single run outcomes with elasticity from continuous probabilities at both the model level and the problem level.
We evaluate numerical and ranking agreement using Spearman’s rank correlation coefficient ($\rho$)\cite{spearman1904} and Kendall’s rank correlation coefficient ($\tau$)\cite{kendall1938}. Prediction error is quantified with Mean Absolute Error (MAE), Root Mean Square Error (RMSE), the Coefficient of Determination ($R^2$), and Symmetric Mean Absolute Percentage Error (sMAPE)~\cite{armstrong1985}.
We further analyze consistency across performance ranges, model families, and size strata to identify when the approximation holds and when it does not.

\textbf{RQ4. How do valence and arousal independently affect correctness and calibration, and what is the evidence for mechanisms?}
Developers’ transient emotions and personalities appear in prompts as interpretable linguistic signals. Concretely, we map each variant’s emotion tag to valence $v\in\{-1,0,1\}$ and arousal $a\in\{-1,0,1\}$ using a fixed dictionary grounded in affective science~\cite{BradleyLang1999ANEW}, and assign it to one of four quadrants (positive--active, positive--calm, negative--active, negative--calm)~\cite{russell1980circumplex,Posner2005CircumplexNeuro}. We also retain fine-grained emotion labels (e.g., focus, confidence, excitement, calm, stress, anxiety, frustration, fatigue) for within-quadrant analysis~\cite{MehrabianRussell1974PAD,BradleyLang1999ANEW}. At the sample level we align performance and confidence and perform grouped aggregations by quadrant and by specific emotion; we compute elasticity and overall indicators, evaluate calibration with ECE, and compare distributional differences with KS. Under controls for length, edit magnitude, term density, and perturbation distance $d$, we quantify the incremental explanatory power of emotion dimensions and use probability dynamics such as confidence trajectories as mechanism evidence.

\subsection{Tasks, Data, and Models}
We use the Python HumanEval benchmark as the initial
dataset. The total size of the enhanced dataset is \textit{14,760}.  Inputs are standardized function signatures and task descriptions. Under \textit{semantic and interface invariance}, we generate semantically equivalent natural language variants using the emotion and personality template system introduced in Section~\ref{subsec:emotion_aware_variant_generation}. Variants are produced at three perturbation distances $d \in \{0.1, 0.2, 0.3\}$. Model outputs are unit tested for pass or fail. 
Variants are generated following the method in Section~\ref{subsec:emotion_aware_variant_generation} and organized as tuples of the original prompt, the perturbation distance, and metadata including emotion and personality tags together with required constraint records, ensuring both interface and semantic invariance. We generate 30 variants per original prompt per distance with random emotion combinations. We follow benchmark terms and do not include adversarial perturbations or sensitive information. Considering the tradeoff between cost and quality, we leverage Deepseek-Chat via API calling to generate the emotion-aware variant.

In the evaluation protocol, all models are assessed on the same test set. To obtain stable continuous measures, we sample 16 independent outputs per prompt at each distance level and apply an identical decoding policy across models. Randomness control is kept consistent throughout by fixing seeds per prompt and distance and reusing them across models.
Quality control enforces \textit{interface and semantic invariance}: function signatures, types, and imports are preserved verbatim, and only the natural language description is rewritten. Template constraints and de duplication remove candidates with semantic drift. For models or samples without probabilities, we adopt \lightapproach{} and report discrete indicators only, without probability imputation.

Our model set spans multiple families and ability ranges to analyze moderation by size and family and to ensure diversity in training paradigms such as instruction tuning and distillation. Representative models include Llama (CodeLlama-34b, CodeLlama-13b, CodeLlama-7b, Llama3.1-8b, Llama-8b distilled, Python-Code-13b), Qwen (Qwen-32b, Qwen-14b, Qwen-7b, Qwen-1.5b, Qwen2.5-Coder-7b), and DeepSeek (DS-Coder-33b, DS-Coder-v2-Lite, DS-Coder-6.7b). For Qwen models, we use variants enhanced via distillation with DeepSeek supervision, except Qwen2.5\mbox{-}Coder\mbox{-}7b, which is evaluated as released.

\textit{Implementation.}
Inference runs on Linux with two A40/L40/L40s GPUs for models up to 30B parameters and one shared H200 GPU for larger models. We use vLLM as the model inference framework. We apply a uniform decoding policy with temperature 0.2 and draw 16 independent samples per prompt at each distance level. When probabilities are available, sequence level log probabilities are aggregated within prompt and normalized with softmax to compute SoftExec ($\text{Acc}_{\text{soft}}$). Prompt variants are produced via the DeepSeek\mbox{-}Coder API following the template design in Section~\ref{subsec:emotion_aware_variant_generation}.

\subsection{Metrics and Statistical Tests}
We quantify prompt stability with a continuous and aggregatable primary measure, complemented by auxiliary indicators for comparison and diagnosis. Robust statistical tests support valid and comparable conclusions.
The primary measure is SoftExec ($\text{Acc}_{\text{soft}}$) with within prompt normalization to obtain continuous correctness, from which we construct elasticity $\text{Elasticity}(p,d)$ and the dataset level curve $\mathcal{E}(d)$. At the dataset level, AUC-E is computed with Simpson’s rule and normalization, aggregating distances into a single scale that captures low probability correctness and uncertainty while enabling direct comparisons across models.
When probabilities are unavailable, we adopt \lightapproach{} proxies, using $\Delta\mathrm{Pass}(d) = \lvert \mathrm{Pass}(0) - \mathrm{Pass}(d) \rvert$ and the \lightapproach{} elasticity $1 - \Delta\mathrm{Pass}(d)$ as interpretable substitutes, consistent with the unified framework in Section~\ref{subsec:sensitivity_evaluation}.
For RQ4, we evaluate calibration with expected calibration error (ECE), which bins predictions by confidence and averages the gap between empirical accuracy and mean confidence, and we compare distribution differences using the Kolmogorov Smirnov statistic (KS), defined as the maximum absolute difference between empirical cumulative distributions.
For statistical testing we use Spearman correlation primarily, with Pearson and Kendall as needed. For group comparisons we apply the Kruskal Wallis test and the Mann Whitney U test, together with bootstrap 95\% confidence intervals~(CI) and Benjamini and Hochberg false discovery rate control. For method consistency in RQ3 we report correlations and error metrics including MAE and RMSE (average absolute and squared error magnitudes), $R^2$ (explained variance, higher is better), and sMAPE (symmetric percentage error, lower is better).

\section{Results and Analysis}
\label{sec:results_analysis}

\subsection{RQ1: Joint Structure of Performance and Prompt Stability}
\label{subsec:rq1_results}

\begin{figure}[]
  \centering
  \scalebox{0.9}{
  \begin{subfigure}{0.48\linewidth}
    \centering
    \includegraphics[width=\linewidth]{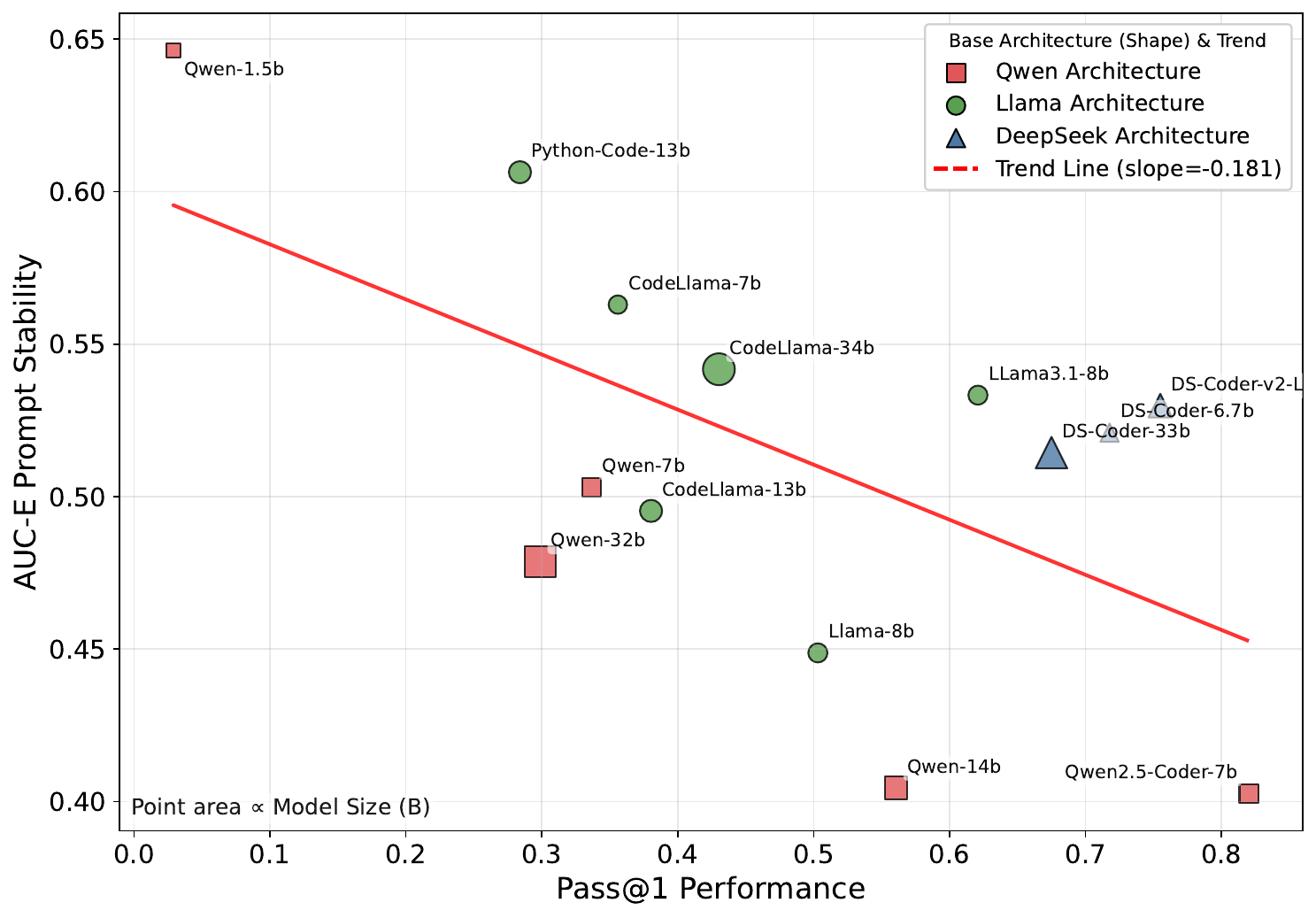}
    \caption{Relationship between performance and stability}
    \label{fig:rq1_main}
  \end{subfigure}\hfill
  \begin{subfigure}{0.48\linewidth}
    \centering
    \includegraphics[width=\linewidth]{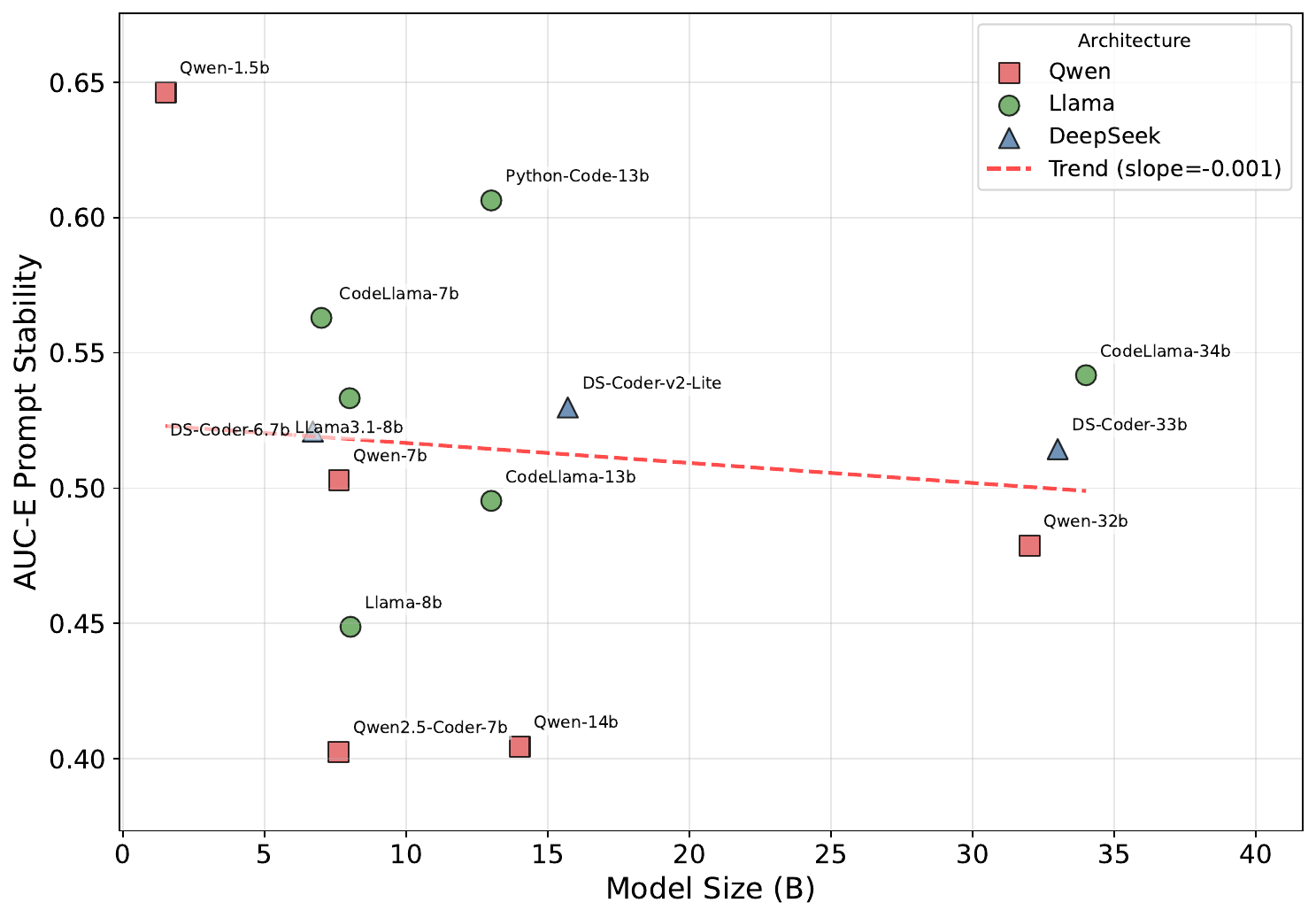}
    \caption{Relationship between AUC-E and parameter scale}
    \label{fig:rq1_size_auc}
  \end{subfigure}}
  \caption{RQ1 comprehensive view at the model level: performance–stability (Pass@1-AUC-E) structure and scale relationship (colored/annotated by architecture).}
\end{figure}

We adopt model-level Pass@1 as the performance indicator and quantify prompt stability through AUC-E (area under the SoftExec-based elasticity curve). Based on the methods in Section~\ref{subsec:sensitivity_evaluation}, we conduct rank correlation analysis at the model level while employing a four-quadrant classification to characterize different models (i.e., combinations of high/low performance and high/low robustness).
Figure~\ref{fig:rq1_main} is based on aggregated model-level data (each model includes core fields such as Pass@1, AUC-E, and parameter scale), covering 14 models; Pass@1 ranges approximately from 0.029–0.820, and AUC-E from 0.404–0.646. Based on the same data, we compute model-level rank correlation, obtaining Spearman $\rho=-0.433$ ($p=0.122$; 95\% CI $[-0.875,0.249]$), which does not reach statistical significance in the current sample, suggesting no unified negative correlation trend. Further dividing models into ``four quadrants'' (high/low performance $\times$ high/low robustness) based on the combination of performance and stability, the corresponding distribution is 3/4/4/3, supporting subsequent stratified analysis by family and scale.
Figure~\ref{fig:rq1_size_auc} is based on the same model-level aggregated data, selecting parameter scale and AUC-E as two core fields to intuitively present the overall relationship between scale and stability, supporting within-family and cross-family comparative analysis.

\textit{Results and Interpretation.} Under the established evaluation protocol, at the current sample size (14 models), we do not observe a statistically significant global negative correlation between Pass@1 and AUC-E (Spearman $\rho=-0.433$, $p=0.122$), suggesting that performance and stability have compatible coexistence space within our sample coverage. Model distribution spans all four quadrants (see counts above), providing diverse selection and configuration paths for practice. This heterogeneity also indicates that architecture and training processes modulate the coupling between performance and stability, making model ``family/scale stratification'' reporting more appropriate than assuming a single universal tradeoff. Two specific examples illustrate this complementary specialization: \textit{Qwen2.5-Coder-7b} has higher Pass@1 ($0.820$) but relatively lower AUC-E ($0.403$); \textit{Python-Code-13b} has lower Pass@1 ($0.284$) but higher AUC-E ($0.606$) (both from the same model-level data).
The scale-stability relationship exhibits non-monotonicity across different families. For example, the Tiny scale (Qwen-1.5B) has higher stability (AUC-E $0.646$); while some medium/large Qwen family models have lower stability; Llama large models (such as CodeLlama-34B) reach medium-high levels (AUC-E $0.542$). This indicates that factors beyond parameter count, such as architecture and distillation, are equally critical; jointly reporting AUC-E with ``scale $\times$ family'' helps identify each model's advantage intervals and suitable scenarios.

\textit{Summary.} 
Within our coverage scope, the performance–stability relationship exhibits diversity rather than a single pattern, with the four‑quadrant structure leaving room for multi‑dimensional selection and customized optimization; scale effects also vary by family. We recommend jointly reporting performance and AUC‑E, stratified by architecture/scale for practical selection. Additionally, supplementary observations of the discrete layer ($\Delta\mathrm{Pass}$) and within‑problem elasticity ($\mathcal{E}(d)$) are consistent with AUC‑E conclusions; in particular, the per‑model summaries include Pass@1 and AUC‑E together with SoftExec means at baseline and at $d\in\{0.1,0.2,0.3\}$, as well as the corresponding elasticity values $E(0.1)$, $E(0.2)$, $E(0.3)$ and their average. These statistics jointly trace the stability curves and their aggregation used in our analysis.

\subsection{RQ2: Distance-Shaped Sensitivity and Scale/Family Moderation}
\label{subsec:rq2_results}

We measure sensitivity across three controlled distances $d\in\{0.1,0.2,0.3\}$. Probability-layer stability employs SoftExec-based elasticity $\mathcal{E}(d)$ and AUC-E; discrete-layer analysis primarily uses $\Delta\text{Pass}(d)$ and its absolute value. The results emphasize scale/family patterns and probability-layer stability.

\begin{figure}[]
  \centering
    \resizebox{0.9\linewidth}{!}{%
    \begin{minipage}{\linewidth}
  \begin{subfigure}{0.45\linewidth}
    \centering
    \includegraphics[width=\linewidth]{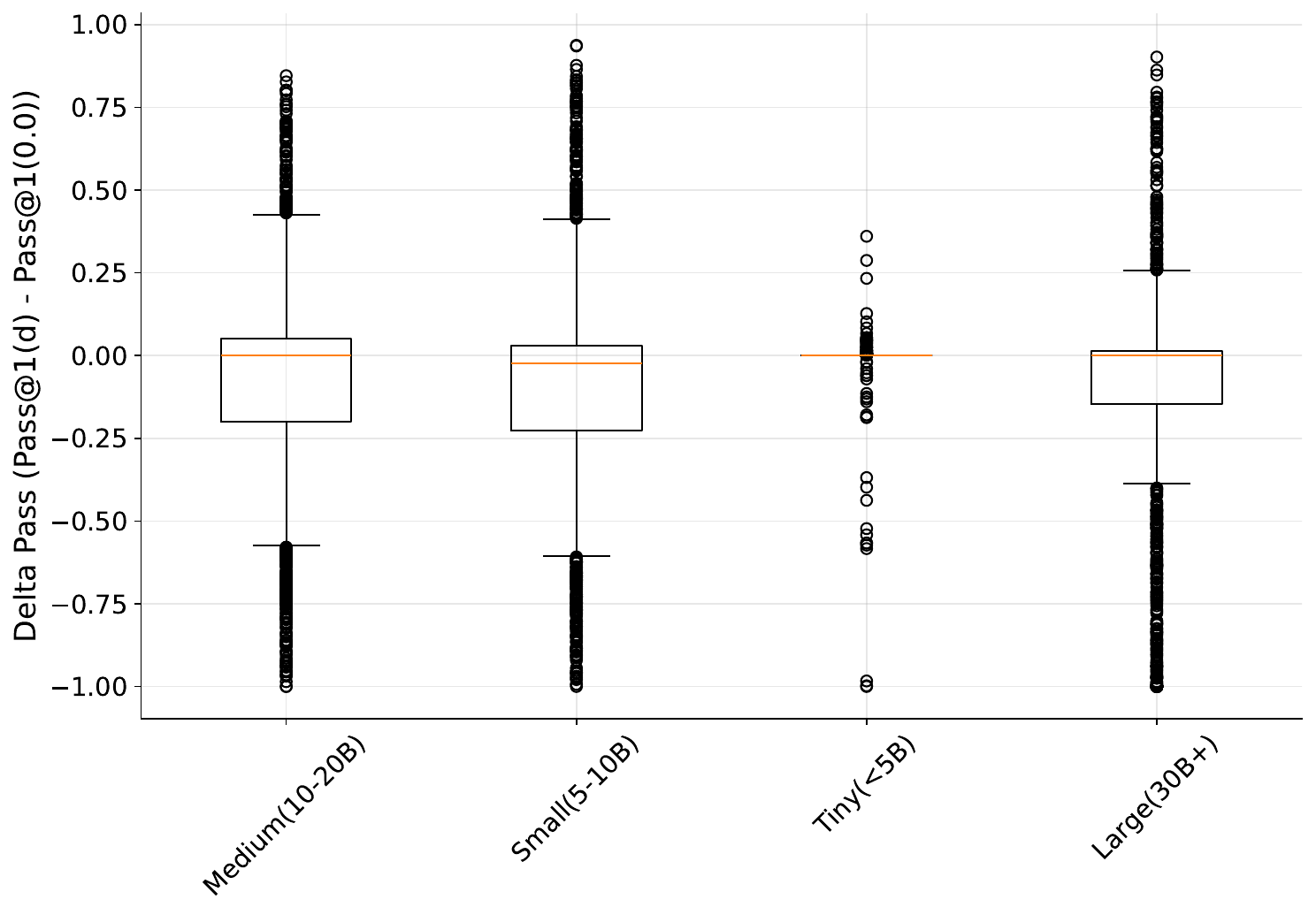}
    \caption{Overall sensitivity by size group ($|\Delta\text{Pass}|$)}
    \label{fig:rq2_size_box}
  \end{subfigure}
  \begin{subfigure}{0.45\linewidth}
    \centering
    \includegraphics[width=\linewidth]{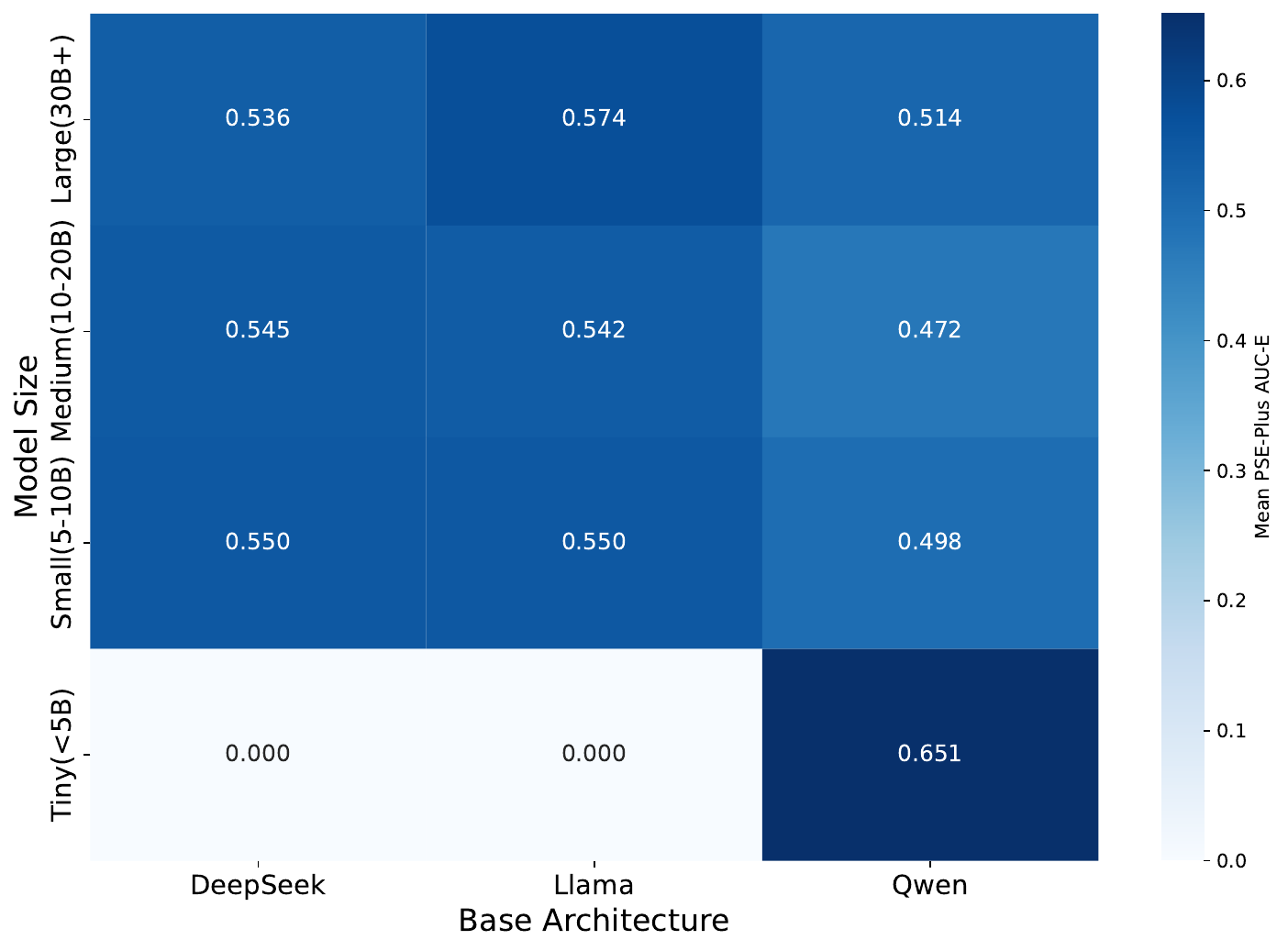}
    \caption{AUC-E by size$\times$architecture}
    \label{fig:rq2_size_arch_heat}
  \end{subfigure}
  \begin{subfigure}{0.45\linewidth}
    \centering
    \includegraphics[width=\linewidth]{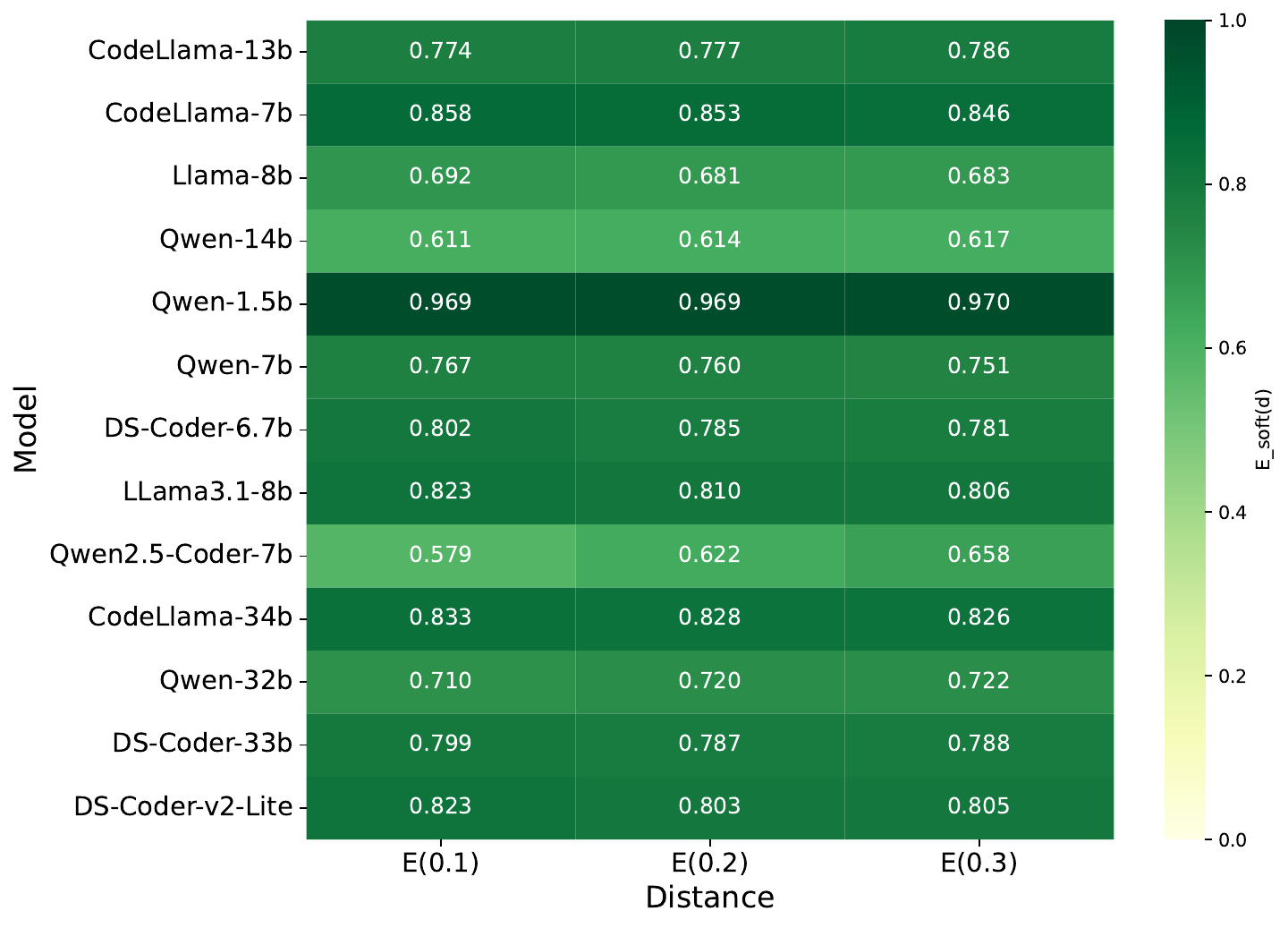}
    \caption{$\mathcal{E}(d)$ elasticity heatmap}
    \label{fig:rq2_esoft_heat}
  \end{subfigure}
  \begin{subfigure}{0.45\linewidth}
    \centering
    \includegraphics[width=\linewidth]{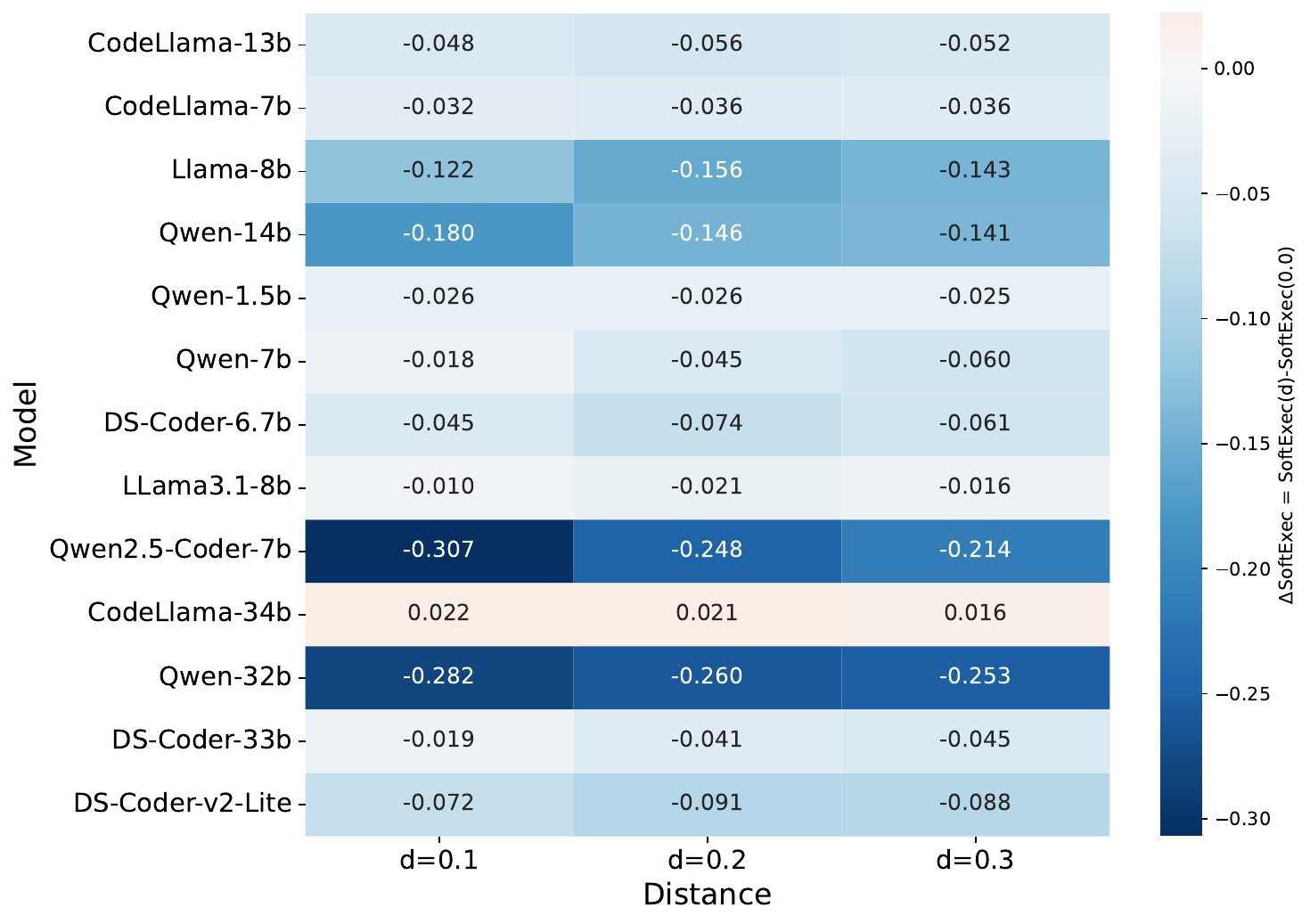}
    \caption{SoftExec changes (relative to baseline)}
    \label{fig:rq2_softexec_heat}
  \end{subfigure}
      \end{minipage}
  }
  \caption{RQ2 comprehensive view: distance-shaped sensitivity and scale/architecture moderation (discrete and probability layers).}
\end{figure}

Figure~\ref{fig:rq2_size_box} aggregates discrete-layer sensitivity across different size groups at each distance, characterized by the mean and variance of $|\Delta\text{Pass}|$, and reports distributional characteristics of each group (including negative proportions, etc.). On the same dataset, we use nonparametric tests to assess main effects of scale, with results showing statistically significant differences (Kruskal–Wallis $H=49.663$, $p=0.000$).
Figure~\ref{fig:rq2_size_arch_heat} groups and aggregates ``scale$\times$base architecture'' by AUC-E at the probability layer, comparing stability levels within families and across scales (including quantification of absolute and relative changes) to characterize the overall robustness landscape of different combinations.
Figure~\ref{fig:rq2_esoft_heat} shows each model's elasticity $\mathcal{E}(d)$ at $d\in\{0.1,0.2,0.3\}$, computed from SoftExec probability-weighted correctness, reflecting stability relative to baseline and its change trajectory with distance.
Figure~\ref{fig:rq2_softexec_heat} characterizes each model's SoftExec changes relative to baseline at different distances, with values closer to 0 indicating closer proximity to baseline, intuitively presenting probability-layer responses under light/medium/heavy rewrites.

\textit{Results and Interpretation.} 
Figure~\ref{fig:rq2_size_box} shows that different size groups show differences in central tendency and variance of $|\Delta\text{Pass}|$: the Tiny group has smaller mean and variance; Large/Small groups have greater variance; the Medium group is intermediate. After cross-model aggregation, the average $|\Delta\text{Pass}|$ values for $d=0.1/0.2/0.3$ are approximately $0.078/0.082/0.078$, showing non-monotonic changes within a narrow range. The significant main effect of scale indicates that scale does have a moderating role under current settings; however, differences still vary by architecture, and we make no causal inferences.
Probability-layer stability (AUC-E) differs across scale$\times$family units (Figure~\ref{fig:rq2_size_box}). For example: the Tiny group has higher averages (mainly driven by 1.5B distilled models); Medium and Large groups show stronger heterogeneity across different families; within the same family, expansion in scale does not show a monotonic trend, suggesting that training data, optimization objectives, and distillation strategies also affect stability.
Most models have $\mathcal{E}(d)$ in higher ranges (approximately $0.75$–$0.85$) across the three distances as shwon in Figure~\ref{fig:rq2_size_arch_heat}; Tiny group models are significantly higher (approximately $0.97$). Family differences are also visible: for example, \textit{Qwen2.5-Coder-7b} has lower elasticity across distances ($\mathcal{E}(d)$ approximately $0.58$–$0.66$), \textit{Qwen-14b} is at the lower-middle boundary ($\mathcal{E}(d)$ approximately $0.61$–$0.62$), while \textit{CodeLlama-7b} drops only about 1.4\% from $d=0.1$ to $d=0.3$. Combined with Figure~\ref{fig:rq2_size_arch_heat}, these phenomena indicate that overall probability-layer performance is relatively stable, but model/family-specific differences exist.
SoftExec changes mostly show mild decline at larger distances as shown in Figure~\ref{fig:rq2_softexec_heat}, consistent with normal fluctuations from stronger rewrites; a few models approach zero or show slight improvement at $d=0.1$, reflecting resilience to light rewrites. Combined with the elasticity perspective, although probability-weighted correctness slightly declines with distance, most models maintain high elasticity due to limited deviation from baseline.

\textit{Summary.} 
Under controlled distances, sensitivity is moderated by scale and family. Probability-layer stability is generally high with architecture-related fluctuations; discrete-layer $|\Delta\text{Pass}|$ shows significant main effects of scale. 

\subsection{RQ3: \lightapproach{} Approximation Effectiveness to \approach{}}
\label{subsec:rq3_results}

We examine whether \lightapproach{} (probability-free) AUC-E approximates \approach{} (probability-aware) AUC-E at the model level and evaluate numerical and ranking consistency, conforming to the settings in Section~\ref{subsec:sensitivity_evaluation}.
Figure~\ref{fig:rq3_compare} compares model-level AUC-E between \lightapproach{} and \approach{}, computing consistency statistics (Pearson/Spearman/Kendall) and error measures (MAE/RMSE/$R^2$) on the same model set, while evaluating ranking consistency. Overall results are: Pearson $r=0.717$, Spearman $\rho=0.723$, Kendall $\tau=0.670$, MAE $=0.040$, RMSE $=0.050$, $R^2=0.41$.

\textit{Results and Interpretation.} \lightapproach{} and \approach{} show moderate to strong monotonic consistency with moderate absolute errors. Overall rankings are basically aligned; a small number of models show more noticeable rank drift (for example, probability-layer scoring favors models with ``good confidence calibration''). In practice, \lightapproach{} can be used for rapid screening, followed by \approach{} for high-fidelity confirmation.

\textit{Summary.} 
Under current settings, \lightapproach{} provides a viable approximation for model-level comparison, though individual models show predictable deviations. 

\subsection{RQ4: Effects of Valence$\times$Arousal on Correctness and Calibration}
\label{subsec:rq4_results}
We analyze correctness sensitivity and calibration under valence$\times$arousal conditions, aggregated by emotion and quadrants, conforming to Section~\ref{sec:method_framework}. The report includes performance–sensitivity structure and \approach{}-based probability-layer confidence diagnostics.

\begin{figure}[]
  \centering
      \resizebox{0.8\linewidth}{!}{%
    \begin{minipage}{\linewidth}
  \begin{subfigure}{0.48\linewidth}
    \includegraphics[width=\linewidth]{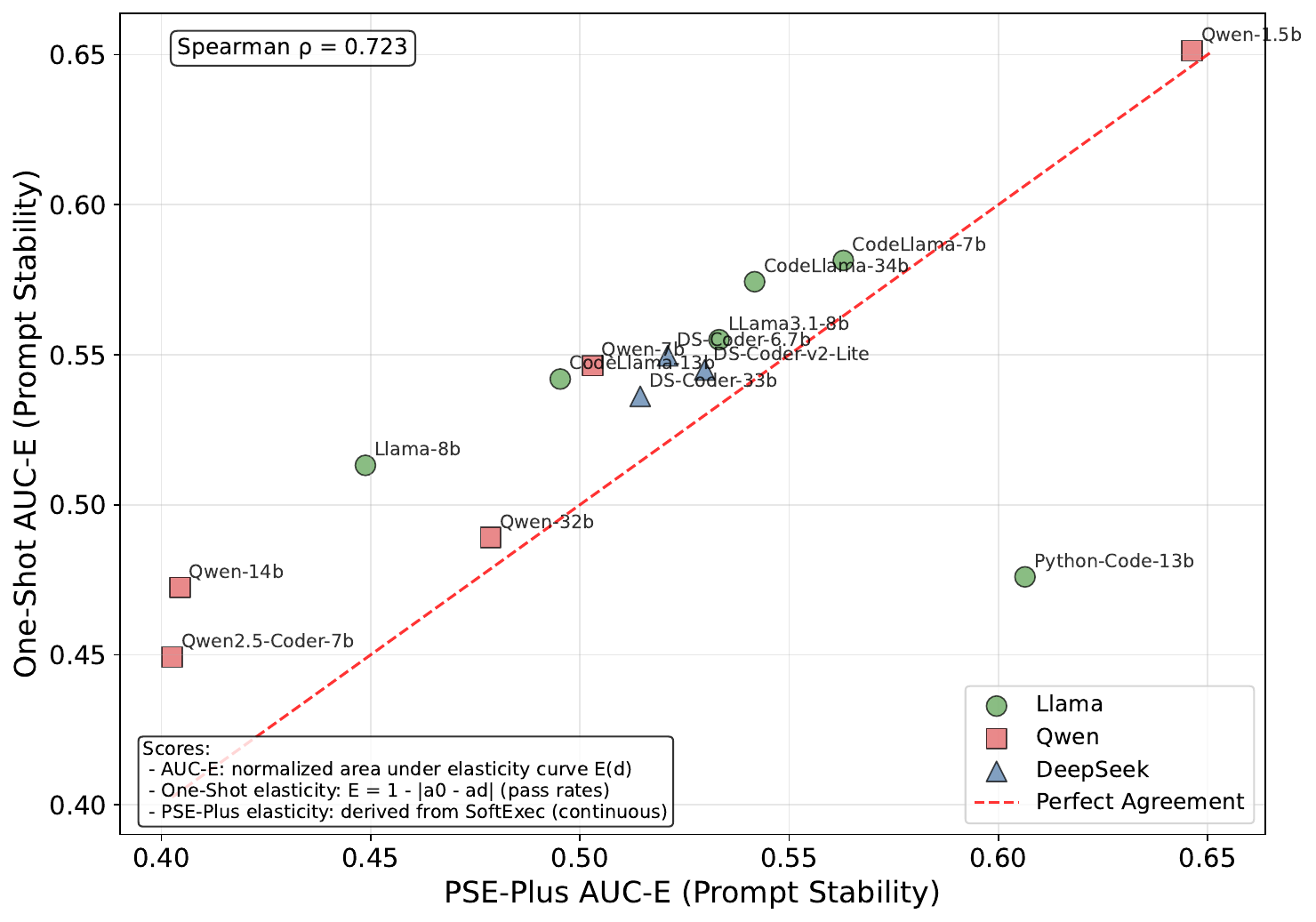}
  \caption{RQ3: Model-level AUC-E: \lightapproach{} vs \approach{}.}
  \label{fig:rq3_compare}
  \end{subfigure}\hfill
    \begin{subfigure}{0.48\linewidth}
    \centering
    \includegraphics[width=\linewidth]{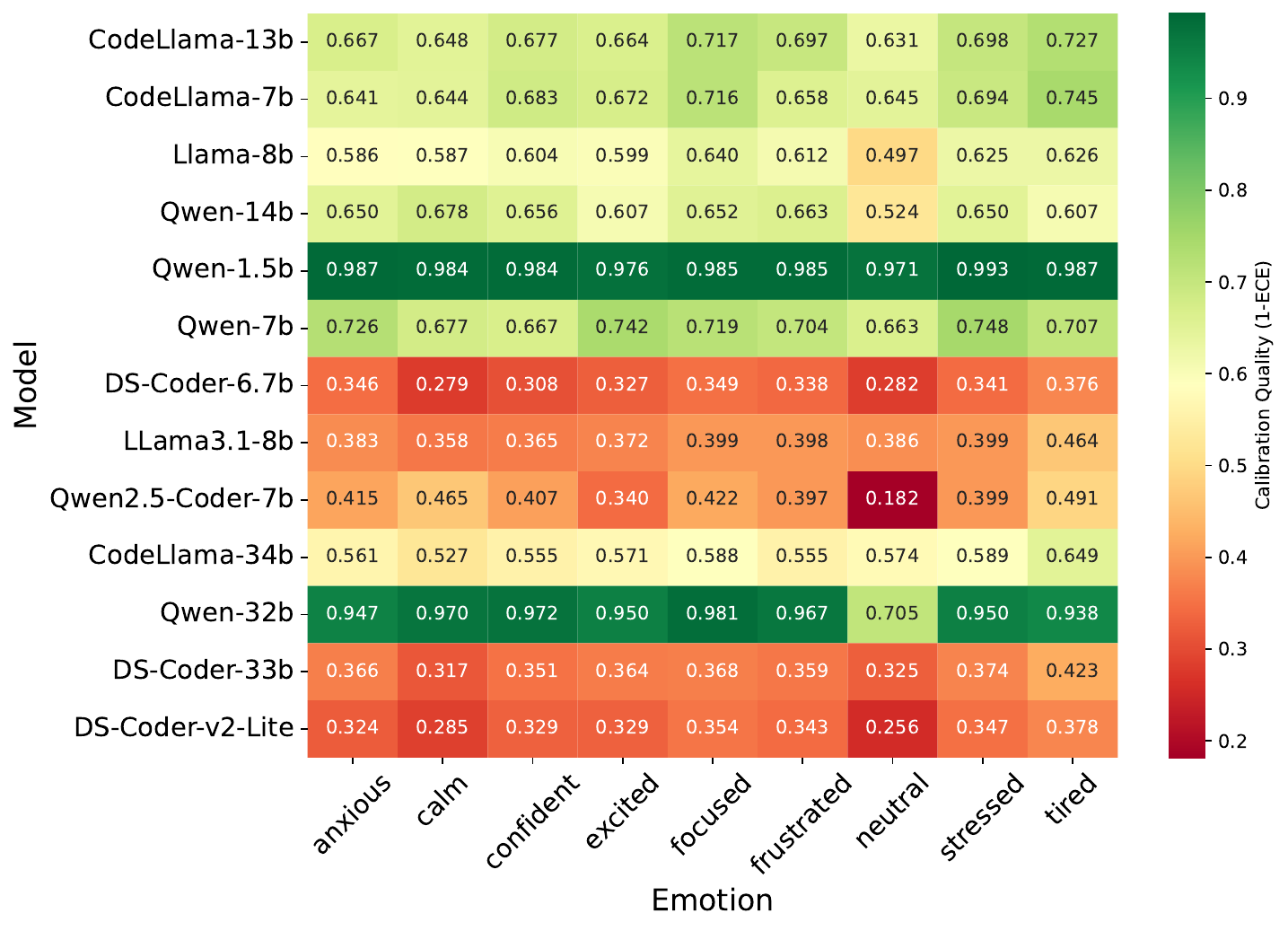}
    \caption{$Model \times emotion$ calibration quality (1-ECE)}
    \label{fig:rq4_calib_heat}
  \end{subfigure}\hfill 
  \begin{subfigure}{0.48\linewidth}
    \centering
    \includegraphics[width=\linewidth]{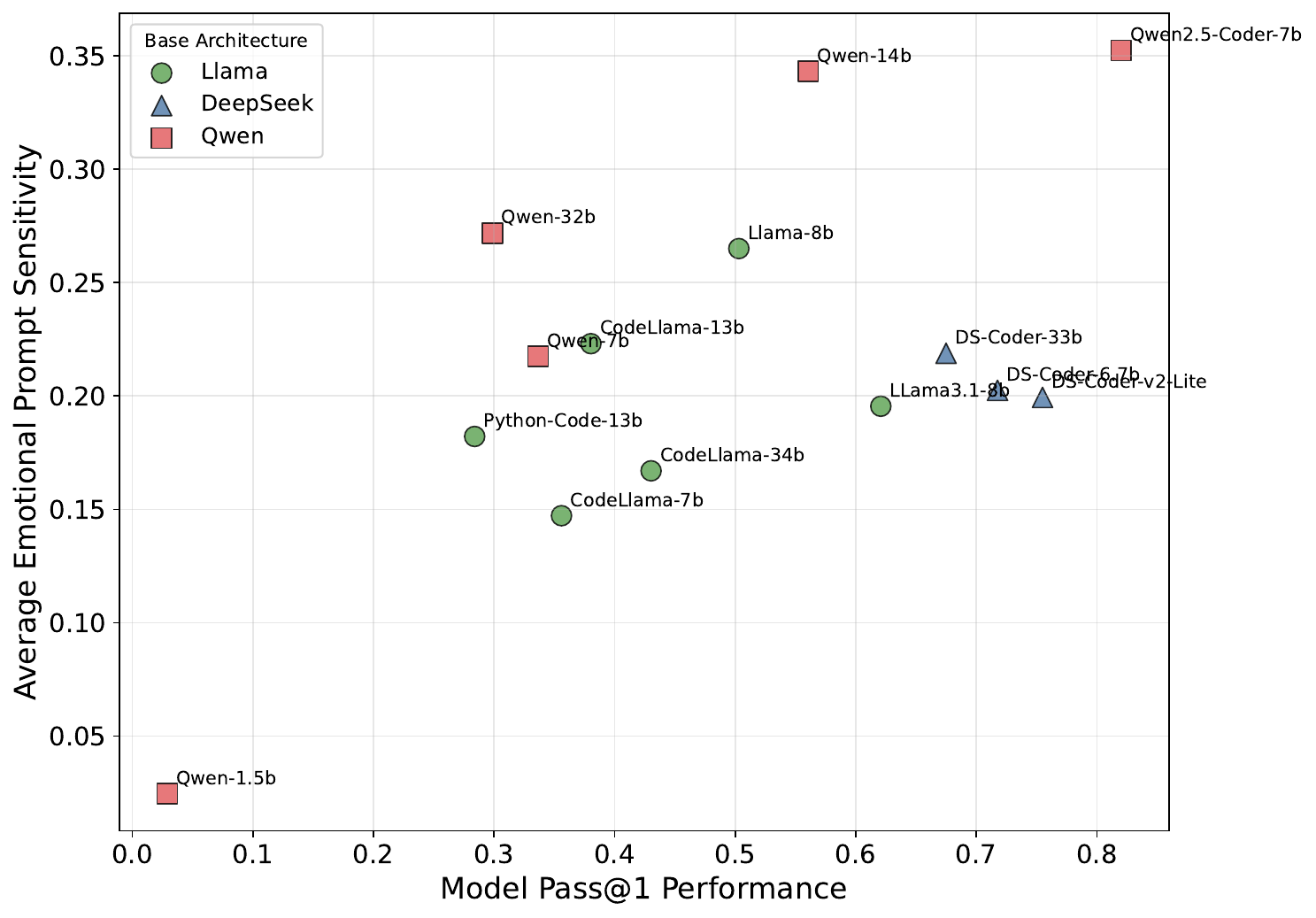}
    \caption{Performance vs average absolute sensitivity}
    \label{fig:rq4_perf_sens}
  \end{subfigure}\hfill
  \begin{subfigure}{0.48\linewidth}
    \centering
    \includegraphics[width=\linewidth]{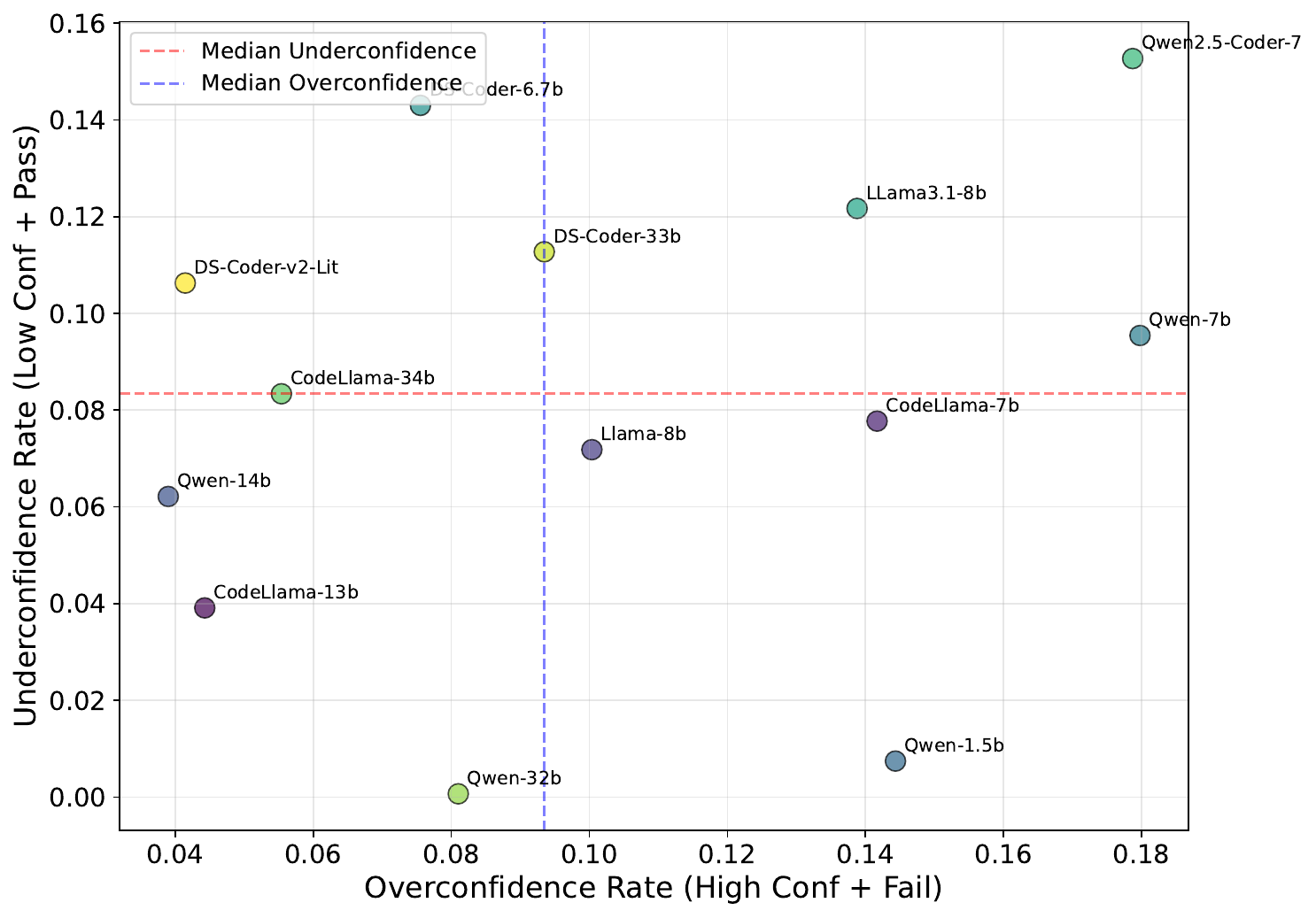}
    \caption{\approach{} confidence bias}
    \label{fig:rq4_bias}
  \end{subfigure}
  \end{minipage}
  }
  \caption{RQ3 (3) Model-level AUC-E: \lightapproach{} vs \approach{}; RQ4 (b-d) comprehensive view: performance–sensitivity, confidence bias, and calibration under emotional conditions.}
\end{figure}

Figure~\ref{fig:rq4_perf_sens} jointly displays model performance (Pass@1) and average absolute sensitivity ($\overline{|\Delta\text{Pass}|}$) under emotional prompting, supplemented with AUC-E, scale, and architecture annotations to facilitate identification of differentiated patterns across families and different scales.
Figure~\ref{fig:rq4_bias} originates from \approach{} confidence analysis based on sample-level probabilities: separately measuring occurrence rates of ``high-confidence failures'' (overconfidence) and ``low-confidence successes'' (underconfidence), aggregated by emotion to model level for comparison; overall calibration uses ECE as background reference to help explain bias profiles across models.
Figure~\ref{fig:rq4_calib_heat} shows $model \times emotion$ calibration quality ($1-ECE$), derived from the same \approach{} confidence analysis aggregated by emotion; probability trajectories and other mechanistic evidence serve as supplementary references to characterize changes in conditional probability distributions.

\textit{Results and Interpretation.} Scatter plots shown in Figure~\ref{fig:rq4_perf_sens} reveal that under emotional prompting, the coupling between performance and discrete sensitivity exhibits heterogeneity. For example, the Tiny group shows lower $|\Delta\text{Pass}|$ at lower Pass@1; some high-performance models show moderate $|\Delta\text{Pass}|$. When aggregated by architecture, average $|\Delta\text{Pass}|$ values are: \textit{Qwen} $=0.242$, \textit{DeepSeek} $=0.207$, \textit{Llama} $=0.197$; differences are mild and model-dependent. Overall, emotion-induced sensitivity varies across models and families, supporting joint reporting of ``performance + AUC-E + sensitivity''.
Models form different ``bias profiles'' in the scatter space (Figure~\ref{fig:rq4_bias} ). When combined with calibration background, model-level ECE ranges from approximately $0.055$ (Qwen-1.5B) to $0.622$ (DS-Coder-6.7B). Joint examination with AUC-E and $|\Delta\text{Pass}|$ can distinguish models with similar correctness but different confidence behaviors.
Warmer colors (higher $1-ECE$) indicate better calibration for that model-emotion pair as shown in Figure~\ref{fig:rq4_calib_heat}. Some models are relatively balanced across emotions; others have weaker calibration on specific emotions. Note that this panel is a $model \times emotion$ summary (without explicit distance axis); distance-related calibration changes are separately characterized by "calibration drift analysis" in the scripts. Combined with performance and stability, this can provide reference for deployment choices in emotion-rich contexts.

\textit{Summary.} 
Under valence$\times$arousal conditions, performance–sensitivity coupling and confidence patterns show cross-model differences. For some models, emotional prompting mildly reshapes correctness and calibration, but effects vary by model and family. Probability-layer diagnostics (ECE, bias, and elasticity) complement discrete indicators, helping make ``calibration-aware'' model selections. Additionally, we use the Kolmogorov–Smirnov (KS) test to compare distributional differences (implementation in analysis scripts and text reports), with qualitative conclusions consistent with ECE/elasticity observations, not presented in figures due to space constraints.

Synthesizing the established perspectives of RQ1–RQ4: (i) performance and stability exhibit structured heterogeneity, providing greater flexibility for model selection and optimization; (ii) distance-driven sensitivity is moderated by scale and family, with probability-layer elasticity generally high and predictable; (iii) \lightapproach{} serves as an effective screening approximation for \approach{}, helping rapidly form reliable rankings under resource constraints, followed by \approach{} confirmation; (iv) under emotional conditions, even with small correctness differences, models remain distinguishable in ``calibration/confidence'' behaviors, providing entry points for deployment strategies and risk management. We recommend jointly reporting Pass@k and AUC-E, with stratified presentation across dimensions of distance, scale, family (and calibration profiles under emotional conditions).

\section{Discussion}
\label{sec:discussion}

\paragraph{\textbf{Model Selection and Performance-Stability Relationships.}}
Our finding that Pass@1 and AUC-E exhibit no statistically significant negative correlation challenges the assumption that stronger models are inherently more robust. The four-quadrant distribution reveals that performance and stability represent distinct optimization objectives that can sometimes be jointly achieved. This provides a practical decision framework: high-performance, high-stability models are most desirable when available, but stable-but-modest models suit reliability-focused production systems, while high-performance-but-sensitive models may excel in controlled research environments. Family-specific patterns (e.g., Llama's balance vs. Qwen's performance orientation) further inform selection based on organizational priorities. Combined with computational cost considerations, this multi-dimensional evaluation enables more informed model procurement and deployment decisions.

\textit{\textbf{The Role of Model Scale and Architecture.}} 
Our observations reveal a non-monotonic relationship between model size and prompt stability in our evaluation. The finding that the Tiny group (e.g., Qwen-1.5B) achieves the highest AUC-E while larger models show greater variance suggests that prompt stability may not follow the same scaling patterns observed for other model capabilities. This could indicate that smaller, well-trained models exhibit certain stability advantages through simpler decision boundaries or reduced overfitting to training prompt distributions. The significant main effect of scale on sensitivity (Kruskal-Wallis $H = 49.663$, $p < 0.001$) combined with family-specific patterns suggests that architectural inductive biases may interact with scale effects. These findings suggest that prompt stability represents a distinct dimension of model behavior that may require explicit optimization during training, rather than emerging automatically with scale.

\textit{\textbf{Emotional and Personality Factors as Robustness Indicators.}}
The systematic effects of valence and arousal on model behavior extend beyond simple performance metrics. High-arousal negative-valence prompts inducing confidence miscalibration in certain models (particularly the Qwen family) suggest that emotional coloring can serve as a practical probe for model brittleness. This phenomenon may reflect training data biases where emotional language correlates with different code quality distributions, causing models to implicitly adjust their confidence based on perceived developer state. The practical implication is that emotion-aware testing can reveal latent instabilities not captured by traditional benchmarks, providing an additional dimension for robustness evaluation.

\textit{\textbf{Prompt Engineering Guidelines.}}
The quantified sensitivity patterns inform prompt engineering best practices. Light perturbations ($d = 0.1$) often preserve performance, suggesting that minor stylistic preferences (e.g., politeness markers) can be accommodated without significant risk. However, heavy changes in expression style ($d = 0.3$) reveal model-specific vulnerabilities, indicating where prompt standardization may be necessary. Development teams can use our emotion templates to stress-test their prompts, identifying phrasings that tend to maintain stability across emotional variations. This is particularly relevant for collaborative development where multiple developers with different communication styles contribute prompts.

\textit{\textbf{Limitations.}}
\label{subsec:limitations}
Our study has several limitations. First, the template-based variant generation may not capture all natural communication patterns, and our emotion/personality dimensions represent a simplified model of human expression. The choice of perturbation distances ($d \in \{0.1, 0.2, 0.3\}$) and use of DeepSeek-Chat for generation may introduce biases affecting sensitivity measurements.
Second, our evaluation focuses on Python code generation using HumanEval, which may not generalize to other programming languages, task domains, or cultural contexts. The emotion and personality templates are designed for English prompts. Our model selection, while covering major families, does not include recent models with different training paradigms.
Third, the AUC-E metric involves design choices that affect interpretation, including equal weighting of perturbation distances and linear aggregation of deviations. The separation of performance and stability as independent dimensions may not fully capture their interaction effects in practical deployment scenarios. Additionally, our execution-based evaluation assumes test suite availability, which may not exist for all real-world coding tasks.

\section{Related Work}

\paragraph{\textbf{Code Generation Evaluation}}
Code generation evaluation has shifted from lexical metrics to execution-based assessment. \citet{chen2021evaluating} introduced HumanEval with pass@k metric for functional correctness, addressing BLEU's limitations \citep{chen2021evaluating}. MBPP uses entry-level tasks \citep{austin2021program}, while APPS evaluates competitive programming \citep{hendrycks2021measuring}. Recent work explores confidence-weighted accuracy: \citet{kadavath2022language} and \citet{jiang2021can} examined how output probabilities correlate with correctness, enabling continuous measures that complement pass@k for analyzing prompt sensitivity \citep{zhang2022opt,liu2023lost}.

\textit{\textbf{Prompt Robustness in LLMs.}}
\citet{ribeiro2018semantically} showed semantically equivalent changes can flip predictions. CheckList formalized invariance testing \citep{ribeiro2020beyond}. In code generation, \citet{he2024prompt} found 40\% performance swings for GPT-3.5 from format changes. Chain-of-thought dramatically improves performance and \citet{kojima2022large} showed ``Let's think step by step'' boosted GPT-3's accuracy from 17\% to 78\%. Template ensembling aggregates multiple phrasings \citep{voronov2024prompt}, while calibration adjusts biases \citep{zhao2021calibrate}, reducing variance by 50\% \citep{liu2023lost}.

\textit{\textbf{Reliability and Calibration Metrics.}}
\citet{liang2022holistic} argue for holistic LLM evaluation covering accuracy, robustness, and calibration. \citet{plaut2025probabilities} found consistent miscalibration across 15 models. ECE quantifies this gap \citep{guo2017calibration}, while variance indicates stability \citep{he2024prompt}. Our work introduces AUC-E (Area Under the Consistency Curve), providing a unified stability metric as perturbations intensify.

\textit{\textbf{Human-Like Factors in Prompts.}}
LLMs reliably adopt personas: \citet{zhang2018personalizing} showed persona-grounding improves dialogue consistency; \citet{wang2025evaluating} demonstrated GPT-4's Big Five trait emulation. Complex personas degrade consistency. \citet{xu2023influence} found prompt emotion affects user perception, while \citet{wu2025personas} showed persona-influenced dialogue strategies. Our work extends this to code generation, examining whether emotional/politeness factors impact correctness---an unexplored domain where functional requirements should theoretically dominate.

\section{Conclusion}
\label{sec:conclusion}

This work introduces \approach{}, the novel measurement framework for evaluating code generation model stability under semantically equivalent prompt variations. While existing benchmarks focus on peak performance, they overlook how developers express identical requirements through diverse phrasings influenced by emotion and personality.
Our evaluation of 14 models reveals three key insights. First, performance and stability show no significant negative correlation, with models distributed across all quadrants, suggesting decoupled optimization objectives. Second, model scale exhibits non-monotonic stability relationships—smaller models can achieve superior robustness. Third, \lightapproach{} strongly approximates \approach{}, enabling rapid assessment for closed-source models.
We make three contributions: (1) Methodologically, principled prompt variant generation through emotion/personality templates that preserve semantics while capturing natural diversity; (2) Technically, SoftExec for probability-aware evaluation and AUC-E for standardized stability quantification; (3) Practically, a dual-pathway system where \lightapproach{} enables screening and \approach{} provides detailed analysis.
Our findings enable informed deployment decision organizations can quantify performance stability tradeoffs, choosing high-stability models for critical systems while accepting sensitivity for research prototypes. Emotion-aware testing reveals latent vulnerabilities beyond traditional benchmarks.

Future work includes understanding architectural stability factors, extending to other tasks, and developing adaptive prompting strategies. Our results suggest prompt stability could join performance, fairness, and safety as a core evaluation dimension. Models that maintain consistent behavior across developer expression diversity will be better positioned for production success than those optimized solely for benchmarks.

\bibliographystyle{ACM-Reference-Format}
\bibliography{reference}

\appendix

\end{document}